\documentclass{PoS}
\usepackage{graphicx}
\usepackage[fleqn]{amsmath}
\usepackage{amssymb}
\newcommand{\be}{\begin{equation}}
\newcommand{\ee}{\end{equation}}
\newcommand{\bea}{\begin{eqnarray}}
\newcommand{\eea}{\end{eqnarray}}
\newcommand{\<}{\langle}
\renewcommand{\>}{\rangle}
\renewcommand{\[}{\langle\!\langle}
\renewcommand{\]}{\rangle\!\rangle}

\newcommand{\nn}{\nonumber}
\title{Scaling study of the relativistic corrections\\ to the static potential}

\ShortTitle{Scaling study of the relativistic corrections to the static potential}

\author{\speaker{Yoshiaki Koma}%
\thanks{Y.K. is partially supported by the Ministry of Education, Science, 
Sports and Culture, Japan,
Grant-in-Aid for Young Scientists (B) (20740149).
M.K. is supported by Japan Society for 
the Promotion of Science (JSPS),
Grant-in-Aid for JSPS Fellows (20$\,\cdot\,$40152).
The authors are also supported by JSPS and DFG under the
Japan-Germany Research Cooperative Program.
The main calculation has been performed on the NEC SX5, SX8 and SX9
at Research Center for Nuclear Physics (RCNP), 
Osaka University, Japan.}\\
Numazu College of Technology\\
E-mail: \email{koma@numazu-ct.ac.jp}}

\author{Miho Koma\\
Numazu College of Technology\\
E-mail: \email{m-koma@numazu-ct.ac.jp}}

\abstract{The relativistic corrections to the static potential, 
i.e. the $O(1/m)$ correction, the $ O(1/m^2) $ spin-dependent and 
momentum-dependent corrections  are investigated in SU(3) lattice gauge theory.
These corrections are relevant ingredients of an effective field theory
for heavy quarkonium called potential nonrelativistic QCD.
Utilizing the multilevel algorithm for the field strength correlator
on the quark-antiquark source, these corrections 
are determined at the distances ranged from 0.25 to 1.2 fm.
A reasonable scaling behavior and long-range nonperturbative contributions
are observed.
}

\FullConference{The XXVII International Symposium on Lattice Field Theory\\
July 26-31, 2009\\
Peking University, Beijing, China}

\begin{document}

\section{Introduction}

A promising way of studying heavy quarkonium systematically
in QCD is to employ an effective field theory called
potential nonrelativistic QCD (pNRQCD)%
~\cite{Brambilla:2000gk,Pineda:2000sz}.
pNRQCD is derived from QCD by integrating out the scale above
the heavy quark mass $m\gg \Lambda_{\rm QCD}$
and the scale of momentum transfer $mv$, where $v$ is quark velocity.

\par
The effective hamiltonian of pNRQCD consists of the nonrelativistic kinetic terms
of a heavy quark and a heavy antiquark with the  inter-quark potentials 
classified in powers of~$1/m$.
Up to $O(1/m^2)$ the effective hamiltonian has the form~\cite{Pineda:2000sz}
\bea
H 
&=&
 \frac{\vec{p}_{1}^{\;2}}{2m_{1}}
+\frac{\vec{p}_{2}^{\;2}}{2m_{2}}
+V^{(0)}(r) 
+\frac{1}{m_{1}}V^{(1,0)}(r)
+\frac{1}{m_{2}}V^{(0,1)}(r) \nn\\
&&
+\frac{1}{m_{1}^2}V^{(2,0)}(r)
+\frac{1}{m_{2}^2}V^{(0,2)}(r)
+\frac{1}{m_{1}m_{2}}V^{(1,1)}(r)
+O(1/m^3) \;,
\label{eqn:pnrqcd-hamiltonian}
\eea
where $m_{1}$ and $m_{2}$ denote the masses 
of quark and antiquark, 
placed at $\vec{r}_{1}$ and $\vec{r}_{2}$, respectively.
$V^{(0)}(r \equiv |\vec{r}|)$, where 
$\vec{r}\equiv\vec{r}_{1}-\vec{r}_{2}$, 
is the static potential.
Note that $m_1$ and $m_2$ can be different.
$V^{(1,0)}(r)=V^{(0,1)}(r)~(\equiv V^{(1)}(r))$
are the corrections at $O(1/m)$.
$V^{(2,0)}(r)$, $V^{(0,2)}(r)$, and 
$V^{(1,1)}(r)$ are the corrections at $O(1/m^2)$, which
contain the leading order 
spin-dependent corrections~\cite{Eichten:1979pu,Gromes:1983pm}
and momentum-dependent
corrections~\cite{Barchielli:1986zs,Barchielli:1988zp}.

\par
The spin-dependent (SD) part of the $O(1/m^2)$ correction 
is conventionally written as
\bea
V_{\rm SD}(r) & =&
\left (
\frac{\vec{s}_{1}\cdot \vec{l}_{1}}{2m_{1}^{2}}
 - \frac{\vec{s}_{2}\cdot \vec{l}_{2}}{2m_{2}^{2}} \right ) 
\left ( \frac{ {V^{(0)\prime}(r)}}{r} 
+2 \frac{ {V_{1}^\prime(r)}}{r} \right )
+\left (
\frac{\vec{s}_{2}\cdot \vec{l}_{1}}{2 m_{1}m_{2}}
-\frac{\vec{s}_{1}\cdot \vec{l}_{2}}{2m_{1}m_{2}}  
\right )\!\! 
\frac{{V_{2}^\prime(r)}}{r}
\nonumber\\ 
&+&
\frac{1}{m_{1}m_{2}}
\left ( \frac{(\vec{s}_{1}\cdot \vec{r})( \vec{s}_{2}\cdot \vec{r})}
{r^{2}} -\frac{\vec{s}_{1}\cdot \vec{s}_{2}}{3} \right ) \! V_{3}(r)
+
\frac{\vec{s}_{1}\cdot \vec{s}_{2}}{3m_{1}m_{2}}V_{4}(r) \; ,
\eea
where $\vec{s}_{1}$ and $\vec{s}_{2}$ denote the spins,
and $\vec{l}_{1}=-\vec{l}_{2}=\vec{l}$ the orbital angular momenta.
Although the original expression of the SD correction in pNRQCD
contains the matching coefficient $c_F$~\cite{Pineda:2000sz}, 
it is assumed to be one here for simplicity.
For the actual application one has to compute the matching coefficient in 
a perturbative or a nonperturbative manner depending on the matching scale
between pNRQCD and QCD.
$V_{1}'(r)$, $V_{2}'(r)$ are responsible for the spin-orbit corrections (fine splitting),
while $V_{3}(r)$, $V_{4}(r)$ are for the spin-spin corrections (hyper-fine splitting).

\par
The spin-independent (SI) part of the $O(1/m^2)$ correction is written as
\bea
V_{\rm SI}(r)
&=&
\frac{1}{m_{1}^2}
\left (\frac{1}{2} \{ {\vec{p}_{1}}^2, V_{p^2}^{(2,0)}(r) \}
+ \frac{1}{r^2} V_{l^{2}}^{(2,0)}(r)
\vec{l}_{1}^{\;2} +V_{r}^{(2,0)}(r) \right )  \nn\\
&&
+
\frac{1}{m_{2}^2}
\left ( \frac{1}{2} \{ {\vec{p}_{2}}^2, V_{p^2}^{(0,2)}(r) \}
+ \frac{1}{r^2} V_{l^{2}}^{(0,2)}(r)
\vec{l}_{1}^{\;2} +V_{r}^{(0,2)}(r) \right )  \nn\\
&&
+\frac{1}{m_{1}m_{2}}
\left (-\frac{1}{2} \{ \vec{p}_{1}\! \cdot \! \vec{p}_{2},
V_{p^{2}}^{(1,1)}(r)\}
-
\frac{1}{2r^2}
V_{l^2}^{(1,1)}(r)
(\vec{l}_{1}\cdot \vec{l}_{2}+
\vec{l}_{2}\cdot\vec{l}_{1})
+V_{r}^{(1,1)}(r)
\right )\; .
\eea
The radial functions  specified by the subscripts $p^2$ and $l^2$,
are related to the momentum-dependent potentials,
$V_{b}(r)$, $V_{c}(r)$, $V_{d}(r)$ and $V_{e}(r)$ defined
in Refs.~\cite{Barchielli:1986zs,Barchielli:1988zp}, by
\bea
&&
V_{{p}^{2}}^{(2,0)} (r)=
V_{{p}^{2}}^{(0,2)} (r)=
V_{d}(r)-\frac{2}{3} V_{e}(r) \; ,\quad
V_{{l}^{2}}^{(2,0)} (r)=
V_{{l}^{2}}^{(0,2)} (r)=
V_{e}(r) \;,\nn\\
&&
V_{{p}^{2}}^{(1,1)} (r)= 
-V_{b}(r) +\frac{2}{3}V_{c}(r) \;, \quad
V_{{l}^{2}}^{(1,1)}(r) = 
-V_{c}(r) \; .
\eea
Once the functional 
forms of these corrections are determined in QCD,
various properties of heavy quarkonium can be investigated systematically 
by solving the Schr\"odinger equation.

\par
Recently, we investigated the $O(1/m)$ 
correction~\cite{Koma:2006si,Koma:2007jq}, 
and the $O(1/m^2)$ spin-dependent~\cite{Koma:2005nq,Koma:2006fw}
and momentum-dependent corrections~\cite{Koma:2007jq}
in SU(3) lattice gauge theory utilizing a new method,
and obtained remarkably clean signals up to distances of around 0.9~fm. 
We observed a certain deviation from the perturbative results 
at these distances.

\par
In this report we present our new results of  the relativistic corrections
at longer distances up to 1.2~fm~\cite{Koma:2008zza}. 
We then examine the scaling behavior 
of the corrections with respect to the change of lattice spacing.

\section{Formulation and Numerical Procedures}

\par
According to pNRQCD, the $O(1/m)$ and $O(1/m^2)$ corrections
can generally be expressed by the {\em matrix elements} and the {\em energy gaps}
appeared in the spectral representation of 
the color-electric and color-magnetic field strength correlators 
(FSCs) on the quark-antiquark source~\cite{Brambilla:2000gk,Pineda:2000sz}.

\par 
Let us explain how to determine the $O(1/m)$ correction as  an example.

\par
We write the eigenstate of the pNRQCD hamiltonian at $O(m^0)$
in the ${\bf 3} \otimes {\bf 3}^{*}$ representation of SU(3) color,
which corresponds to the {\em static} quark-antiquark state,
as $| n \> \equiv | n; \vec{r}_{1},\vec{r}_{2} \>$.
Then, the color-electric FSC,
i.e.~the correlator of two color-electric field strength operators
$E_{i} =F_{4i}$ ($i=1,2,3$), attached to a quark
at $\vec{r}_{1}$ and separated $t=t_{1}-t_{2}$ in the time direction, 
takes the form
\bea
 C(r,t) 
= \sum_{n=1}^{\infty} \< 0 | gE_{i}(\vec{r}_{1}) |n\>
\< n| gE_{j}(\vec{r}_{1})
| 0\> e^{-(\Delta E_{n0}(r))t}\;,
\label{eqn:correlator-infinite}
\eea
where $\Delta E_{n0}(r) \equiv E_{n}(r)-E_{0}(r)$ denotes
the energy gap and $E_{0}(r) =V^{(0)}(r)$.
The $O(1/m)$ correction is expressed with the matrix elements
$\<0|g E_{i}(\vec{r}_{1}) |n\>$ and the energy gap $\Delta E_{n0}(r)$
as
\bea
V^{(1)}(r) = - \frac{1}{2}\delta^{ij}\sum_{n = 1}^{\infty}
\frac{\< 0 | g E_{i} (\vec{r}_{1})| n\> \< n | g E_{j} (\vec{r}_{1})| 0\>}
{(\Delta E_{n0}(r) )^2} \; .
\label{eqn:v1-spectralrep}
\eea
Thus, once the matrix elements and the energy gaps
are known from the behavior of FSCs,
one can compute the correction.

\par
We work in Euclidean space in four dimensions on a
hypercubic lattice with lattice volume $V=L^3 T$ and 
lattice spacing $a$, where we impose
periodic boundary conditions in all directions.
We use the Polyakov loop correlation function (PLCF, 
a pair of Polyakov loops $P$ separated by a distance $r$)
as the quark-antiquark source
and evaluate the color-electric 
FSCs on the PLCF,
\bea
C(r,t) =\[ g E_{i} (\vec{r}_{1},t_{1}) g E_{j} (\vec{r}_{1},t_{2}) \]_{c}
=
\[ g E_{i} (\vec{r}_{1},t_{1}) g E_{j} (\vec{r}_{1},t_{2}) \]
\!- \! \[ g E_{i} (\vec{r}_{1})\]  \[ g E_{j} (\vec{r}_{1})\] \;,
\label{eqn:correlator}
\eea
using the multi-level algorithm~\cite{Koma:2006si,Koma:2006fw},
where the double bracket represents the ratio of expectation values
$\[\cdots \] =  \< \cdots \>_{PP^{*}} /  \< P P^{*}(r)\>$, while
$\< \cdots \>_{PP^{*}}$ means that 
the color-electric field is connected to 
the Polyakov loop in a gauge invariant way.
The subtracted term on the r.h.s. of Eq.~\eqref{eqn:correlator}
can be nonzero as the color-electric field is even 
under $\mathit{CP}$ transformations.
The spectral representation of Eq.~\eqref{eqn:correlator} 
derived with transfer matrix theory reads~\cite{Koma:2006fw}
\bea
C(r,t) 
=
2 \!\! \sum_{n = 1}^{\infty} 
\< 0  | g E_{i}(\vec{r}_{1}) | n \> 
\< n | g E_{j} (\vec{r}_{1}) | 0 \>
e^{- (\Delta E_{n0})T/2} 
\cosh ((\Delta E_{n0})(\frac{T}{2} -t)) \! 
+ \!O(e^{-(\Delta E_{10})T})\;  ,
\label{eqn:correlator-spectralrep}
\eea
where the last term represents terms 
with exponential factors equal to or smaller than
$\exp (-(\Delta E_{10})T)$, 
which are negligible
for a reasonably large $T$.
We evaluate Eq.~\eqref{eqn:correlator} via Monte Carlo 
simulations and determine the matrix element
$\< 0 | g E_{i}(\vec{r}_{1}) |n \>\< n | g E_{j}(\vec{r}_{1}) |0 \>$
and the energy gap $\Delta E_{n0}$ 
in Eq.~\eqref{eqn:correlator-spectralrep} by fitting,
both of which are finally inserted into 
the definitions of the corrections, such as
Eq.~\eqref{eqn:v1-spectralrep}.
Eq.~\eqref{eqn:correlator-spectralrep} is reduced to
the form like Eq.~\eqref{eqn:correlator-infinite}
in  the infinite volume limit $T\to \infty$.

\par
We define 
the lattice color-electric field operator,
$g a^2 E_{i}(s)$, from the traceless part of
$[ U_{4i}(s)-U^{\dagger}_{4i}(s)]/(2i)$
with two-leaf modification (an average of $F_{4i}(s)$
and $F_{4i}(s-\hat{i})$), where $U_{\mu\nu}(s)$ is a
plaquette variable defined on the site $s$.
We multiply the Huntley-Michael
factor~\cite{Huntley:1986de} on the PLCF,
$Z_{E}$~\cite{Koma:2006fw},
to the lattice color-electric field 
to cancel the self energies at least at $O(g^2)$.

\par
Other relativistic corrections can be investigated similarly.
See Refs.~\cite{Koma:2007jq,Koma:2006fw} 
for definitions and further technical details.

\section{Numerical results}

We carry out simulations using
the standard Wilson gauge action in SU(3) lattice gauge theory.
Simulation parameters are summarized in Table~\ref{tbl:para}.

\newcommand{\lw}[1]{\smash{\lower2.0ex\hbox{#1}}}
\begin{table}[t]
\caption{Simulation parameters used in this study.
$N_{\mathrm{tsl}}$ is the number of time slices in a sublattice
and $N_{\mathrm{iupd}}$ the number of internal update within a
sublattice, both are parameters for the multilevel 
algorithm~\cite{Koma:2006fw}. 
The lattice spacing $a$ is set from the Sommer scale
$r_{0}=0.5$~fm.
}\vspace*{0.2cm}
\centering
\begin{tabular}{ccc|ccc|ccc}
    \hline
    \lw{$\beta=6/g^2$}& \lw{$a$ [fm] }&
   \lw{ $N_{\mathrm{tsl}}$}  &
   \multicolumn{3}{|c}{spin-dependent corrections}&
   \multicolumn{3}{|c}{spin-independent corrections}\\
    &&& $(L/a)^3 (T/a)$& $N_{\mathrm{iupd}}$ & 
    $N_{\mathrm{conf}}$
    & $(L/a)^3 (T/a)$& $N_{\mathrm{iupd}}$ & 
    $N_{\mathrm{conf}}$\\
    \hline
    5.85 & 0.123 &   3 & $24^{4}$ & 50000 & 77 &$ 24^{4}$ & 50000 & 
    133 \\
    6.00 & 0.093 &   4 & $20^{3}40$ & 7000 & 33 &$ 24^{3}32$ &50000& 100\\
    6.20 & 0.068 &   5 & 
      $24^{3}30$ & 10000 & 33 & $30^{3}40$ & 50000 &33 \\
    6.30 & 0.059 &   6 & $24^{4}$ & 6000 & 39 &
\\
    \hline 
\end{tabular}
\label{tbl:para}
\end{table}

\newcommand{\figw}{7.7cm}
\begin{figure}[hbt]
\includegraphics[width=\figw]{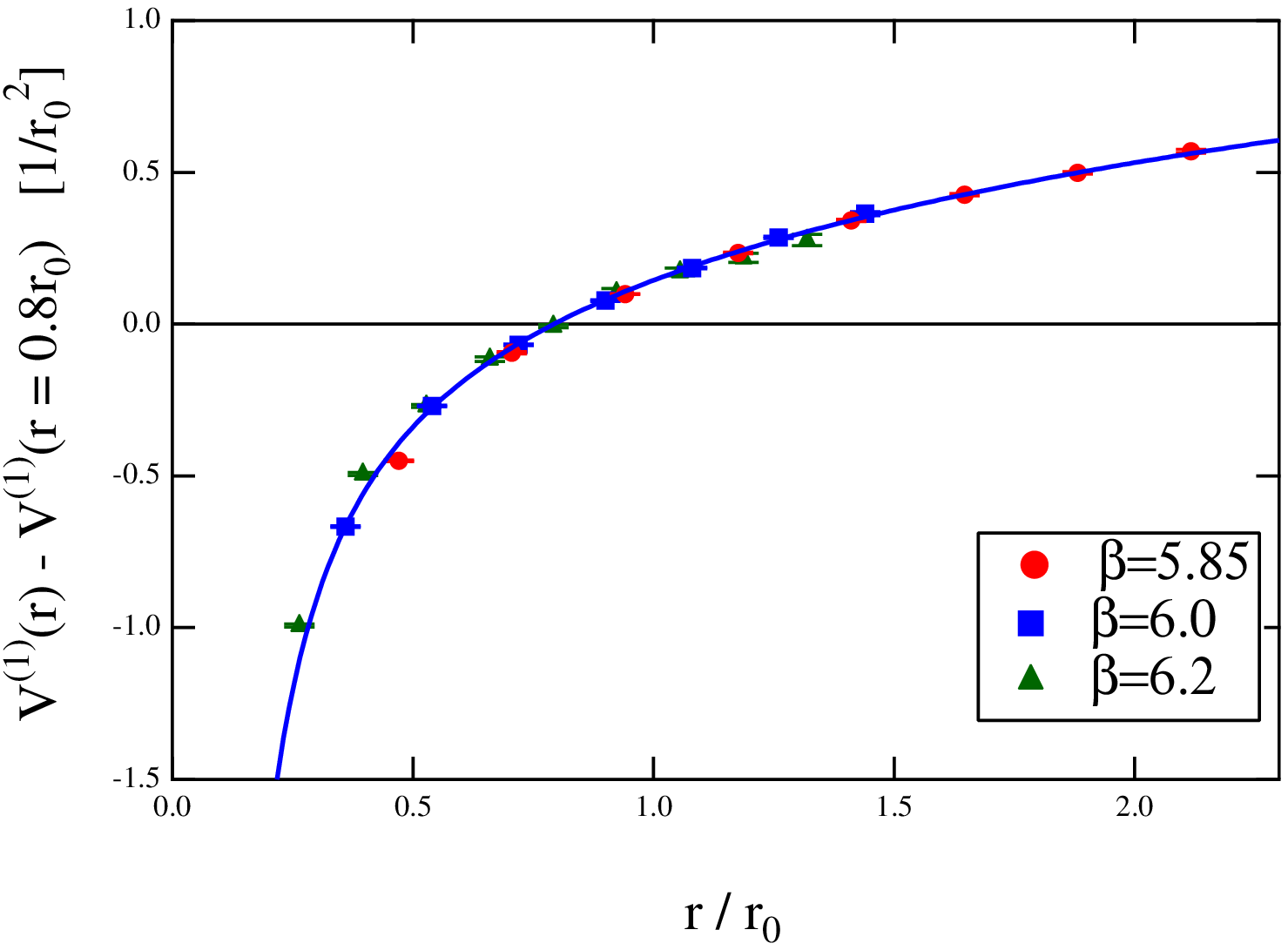}
\includegraphics[width=\figw]{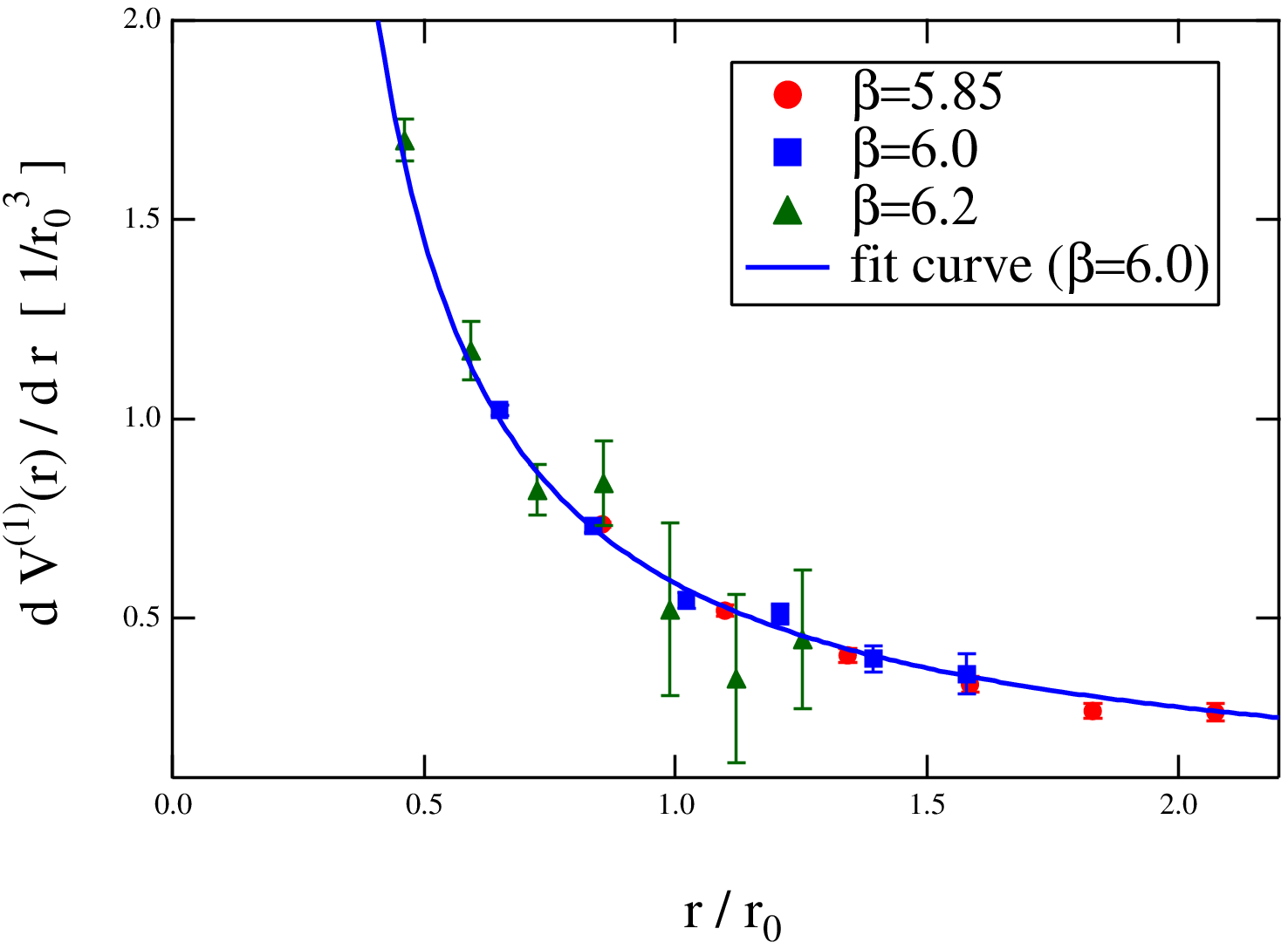}
\caption{The $O(1/m)$ correction, $V^{(1)}(r)$, normalized at $r=0.8\, r_{0}$
(left) and its derivative with respect to $r$ (right) in units of $r_{0}$.
The solid line is the fitting curve for the data at $\beta=6.0$
with the functional form $V_{\rm ln}(r)=-A/r^{2}+B\ln r+C$ and 
$dV_{\rm ln}(r)/dr$,
where the data at $r/a=2$ is not taken into
account in the fit.
}
\label{fig:pot-v1m}
\end{figure}

\subsection{The $O(1/m)$ correction}

In Fig.~\ref{fig:pot-v1m}, we show the $O(1/m)$ correction, $V^{(1)}(r)$,
normalized at  $r=0.8\, r_{0}$ together with the fitting curve
$V_{\rm ln}(r)=-A/r^{2}+B\ln r+C$.
The first term is motivated by 
perturbation theory at $O(\alpha_{s}^{2})$, which provides
$V_{\rm pert}(r) = - \frac{C_{F}C_{A}\alpha_{s}^{2}}{4r^{2}}$,
where $C_{F}$ and $C_{A}$ are the Casimir charges of the fundamental 
and the adjoint representations, respectively, and 
$\alpha_{s}=g^2/(4\pi)$ the strong coupling~\cite{Brambilla:2000gk}.
We measure $V^{(1)}(r)$ up to $r=2.35 \, r_{0} =1.2$~fm. 
$V^{(1)}(r)$ for different $\beta$ values
show a reasonable scaling behavior, except for the data at $r/a=2$,
which are always suffered from a larger discretization error than the 
data at $r/a\geq 3$. 
We find that the $O(1/m)$ correction contains the radial behavior which
cannot be explained by perturbation theory at $O(\alpha_{s}^{2})$.
The new data set at $r  \gtrsim 1.5 \, r_{0}$ 
suggests that the long distance behavior 
is described by a logarithmic function~\cite{Koma:2008zza,PerezNadal:2008vm}.

\begin{figure}[tb]
\includegraphics[width=\figw]{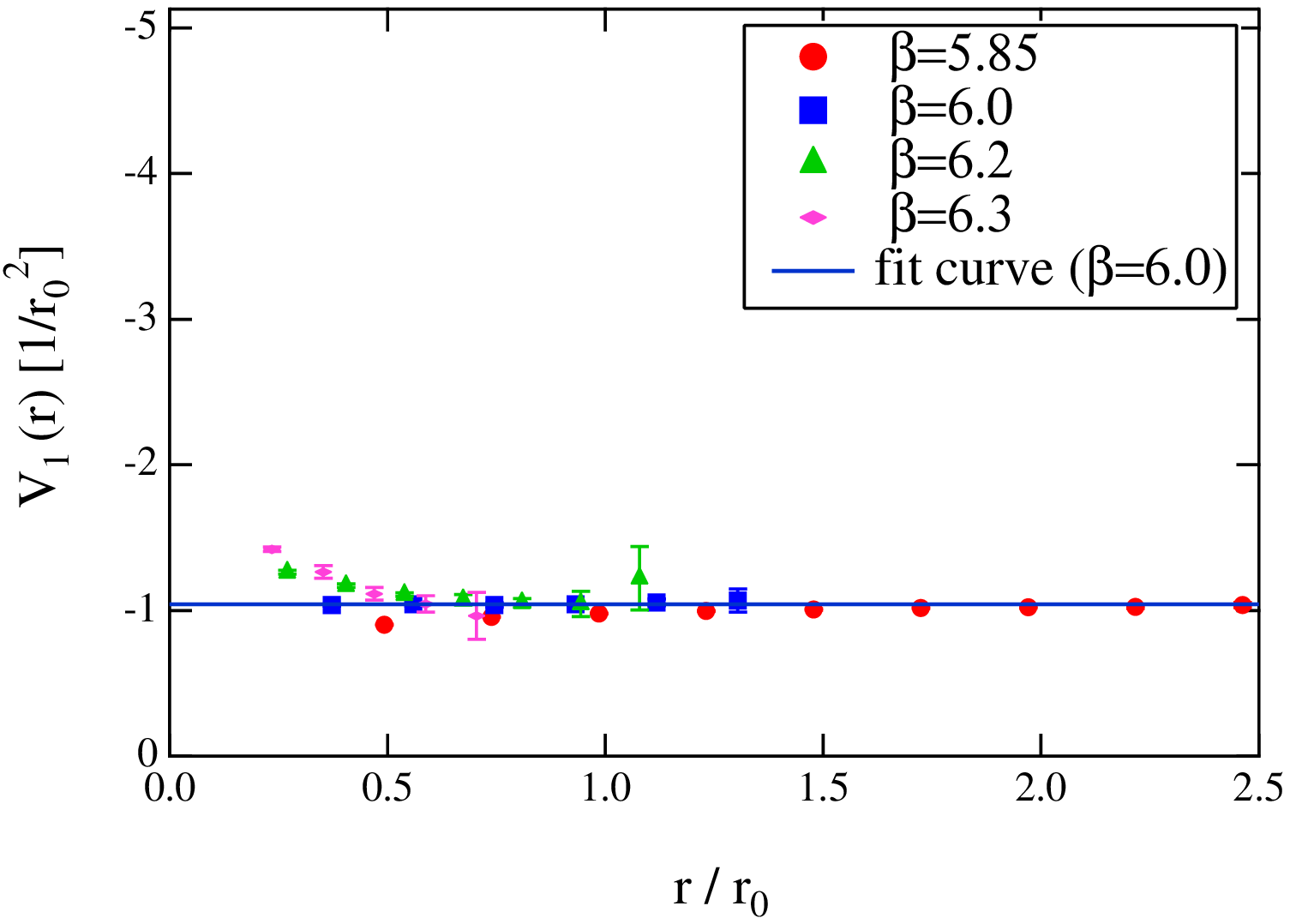}  
\includegraphics[width=\figw]{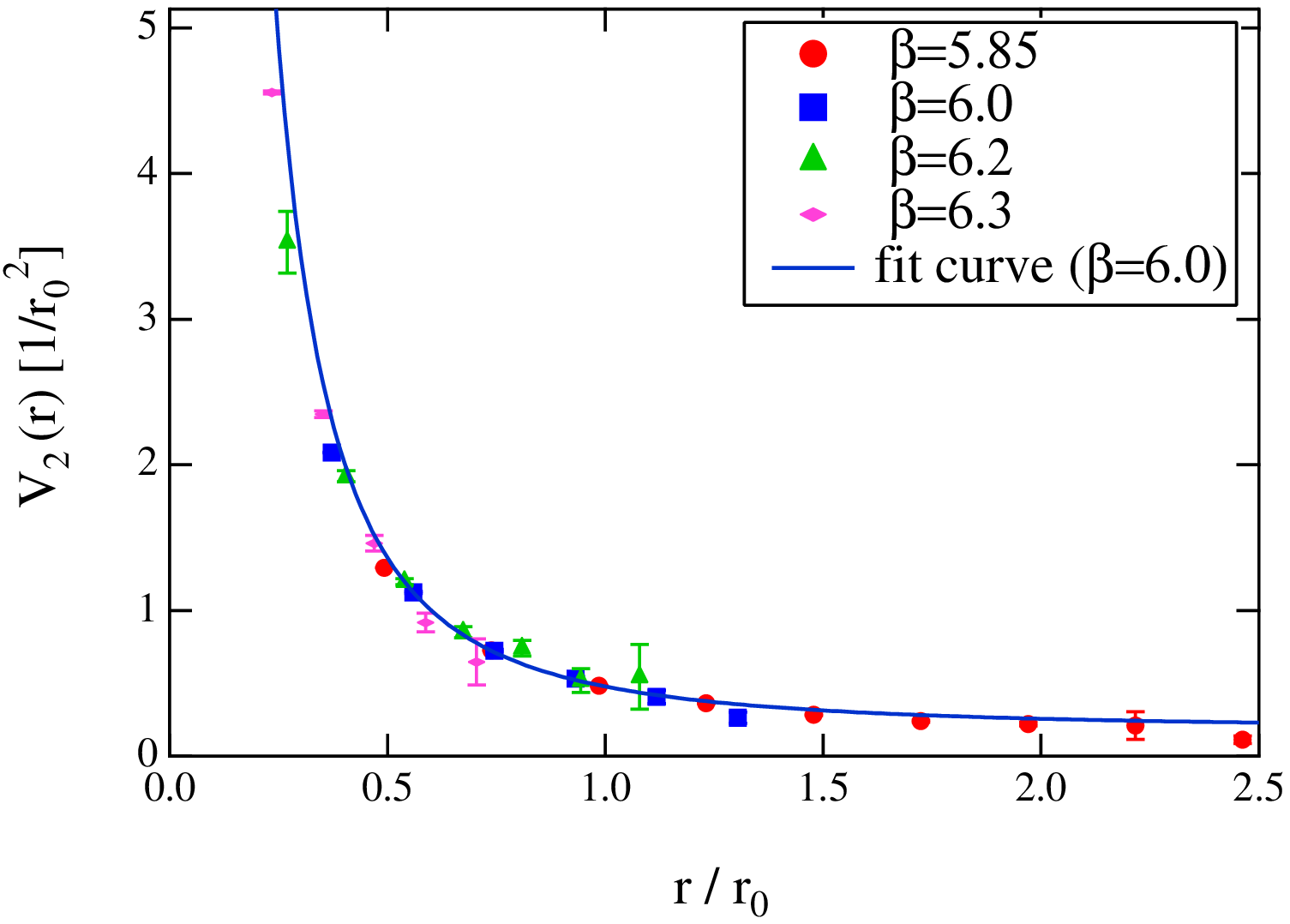}\\
\includegraphics[width=\figw]{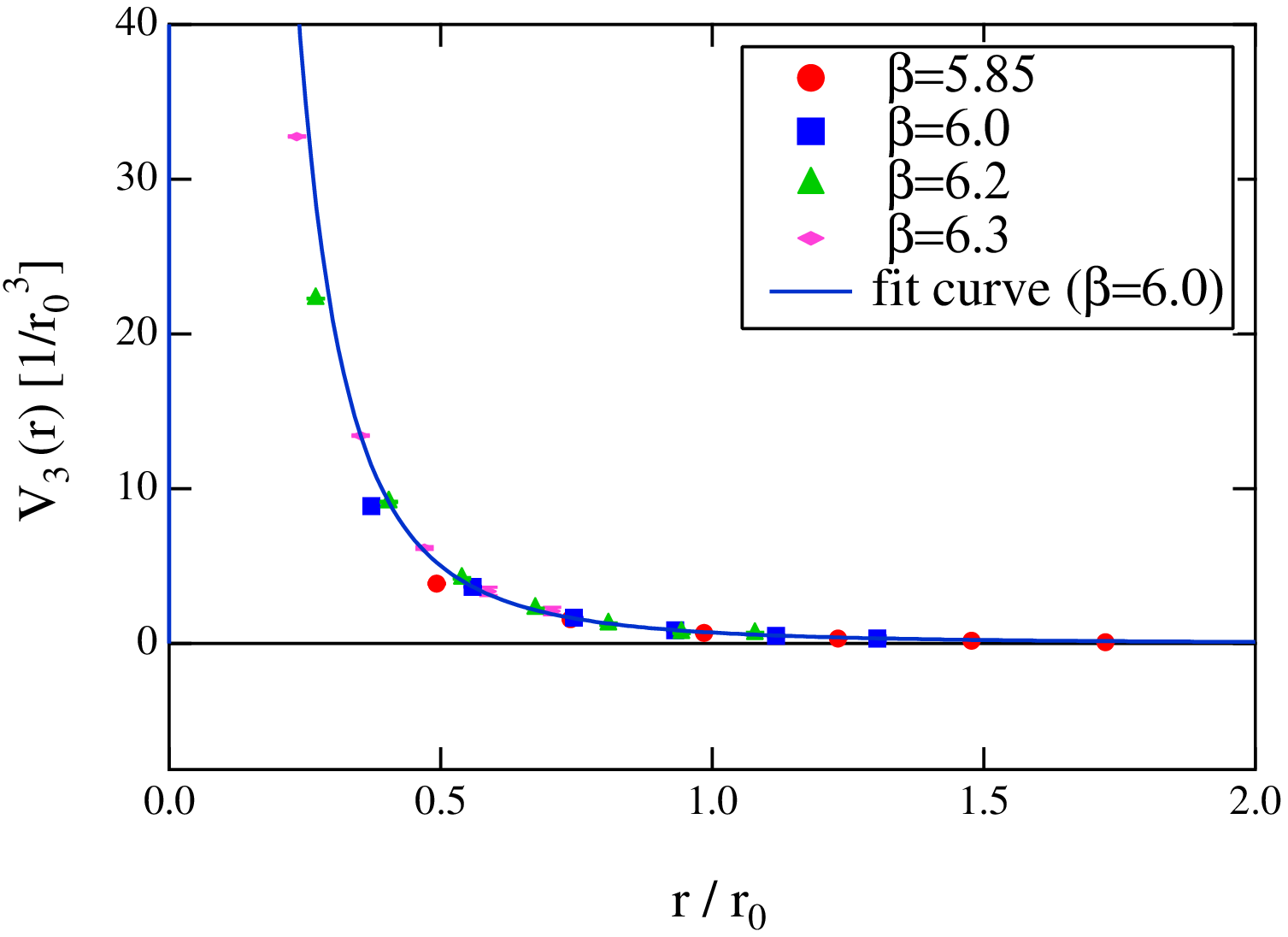}  
\includegraphics[width=\figw]{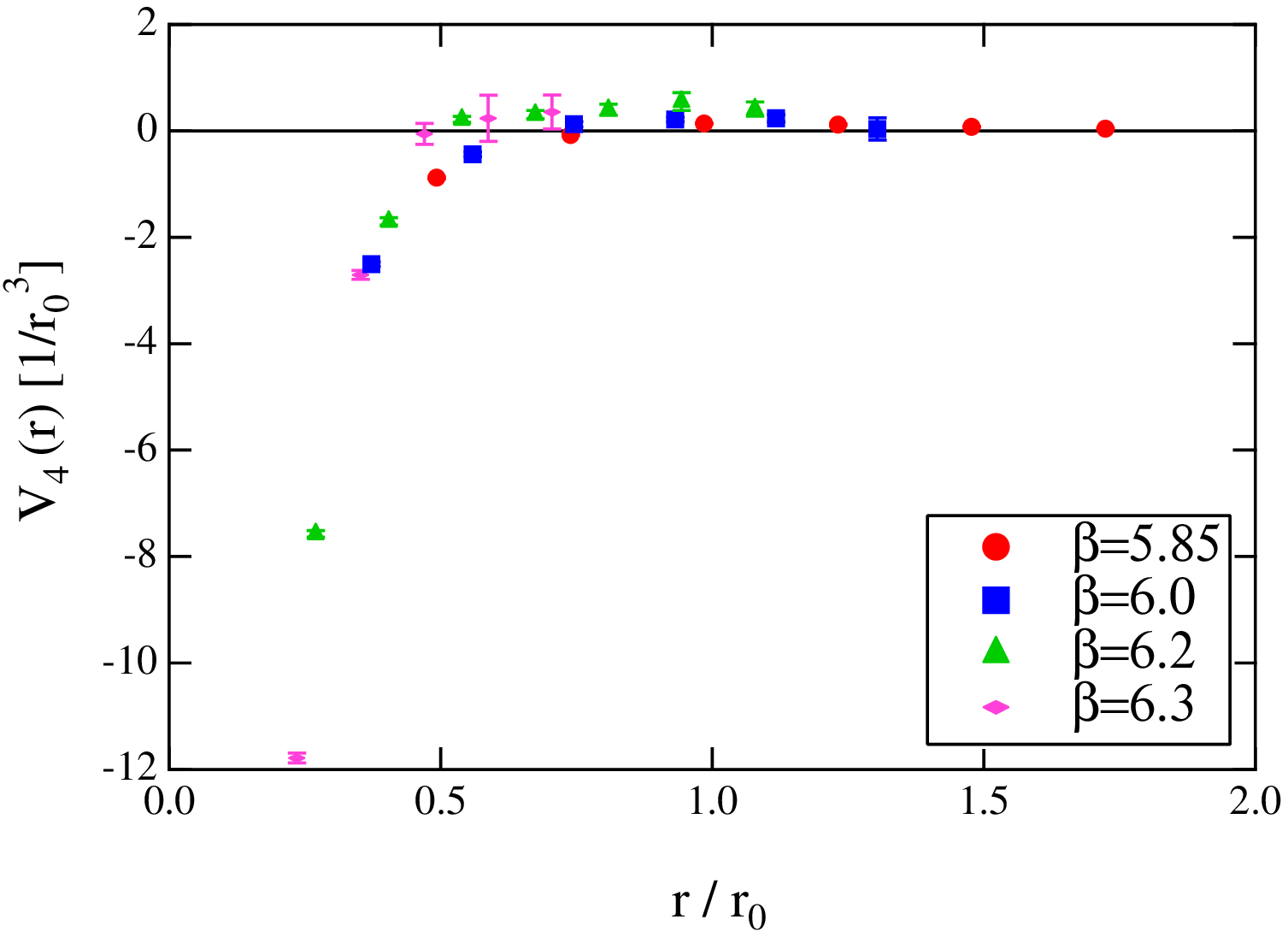}  
\caption{The spin-dependent corrections $V'_{1}(r)$, $V'_{2}(r)$, $V_{3}(r)$
and $V_{4}(r)$ in units of $r_{0}$.}
\label{fig:pot-spin}
\end{figure}
\begin{figure}[tb]
\includegraphics[width=\figw]{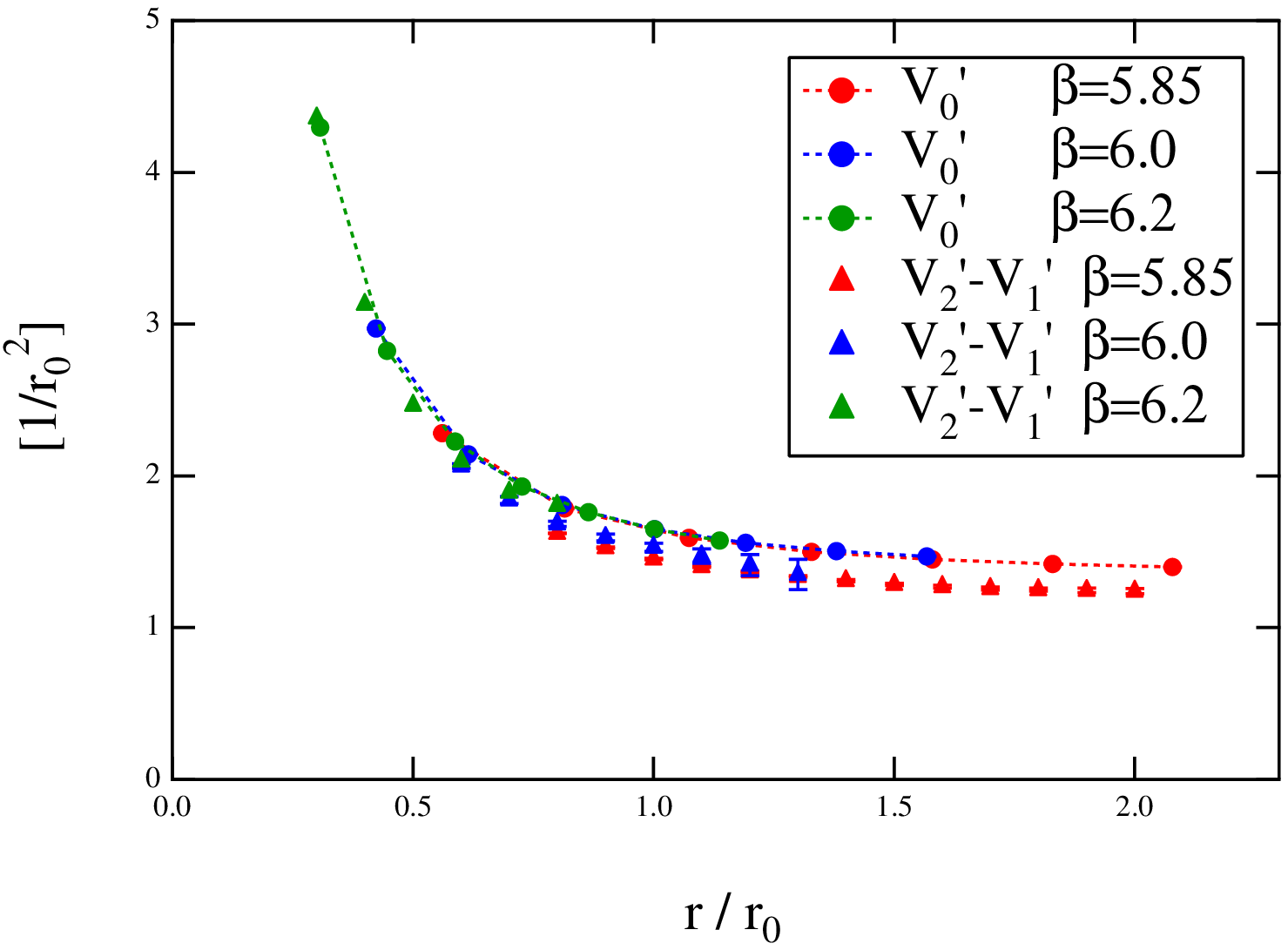}
\includegraphics[width=\figw]{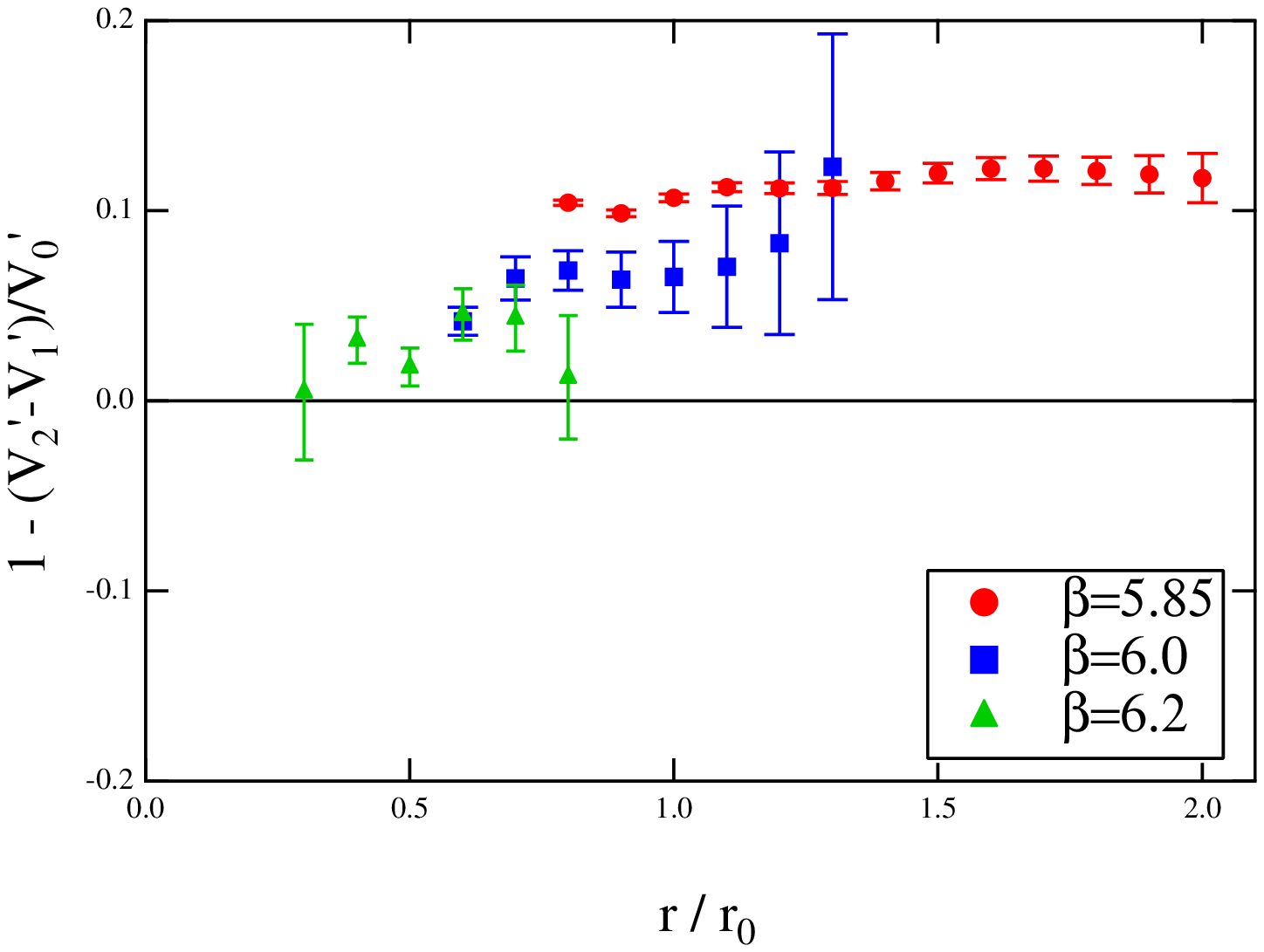}  
\caption{Test of the Gromes relation $V^{(0)\prime}(r)=V_{2}'(r)-V_{1}'(r)$.
Comparison between the force $V^{(0)\prime}$ and 
the difference of the spin-orbit corrections 
$V_{2}'-V_{1}'$ (left), and the relative deviation from
the Gromes relation 
$1-(V'_{2}-V'_{1})/V^{(0)\prime}$ (right).}
\label{fig:Gromes}
\end{figure}

\begin{figure}[tb]
\includegraphics[width=\figw]{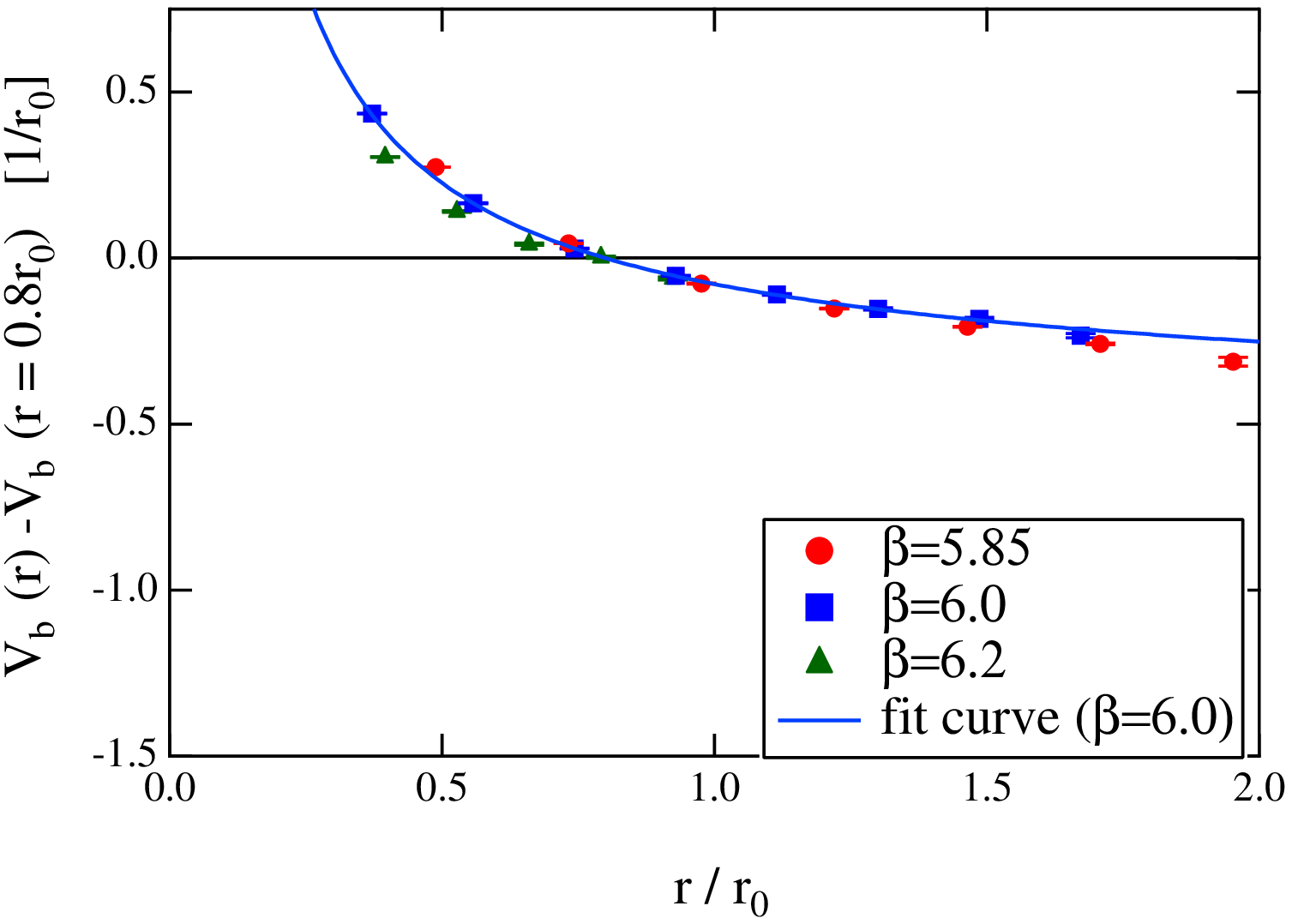}  
\includegraphics[width=\figw]{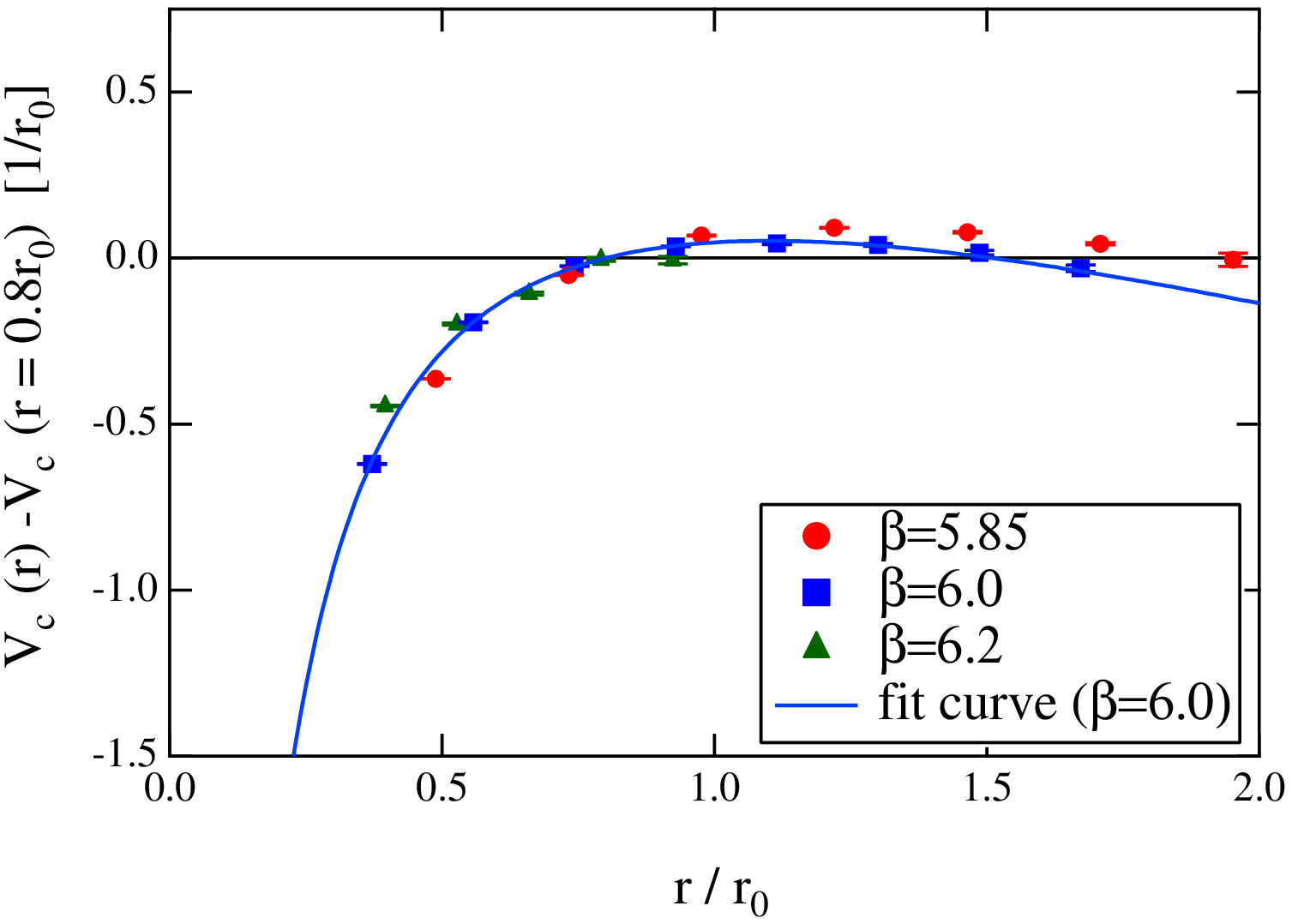}\\
\includegraphics[width=\figw]{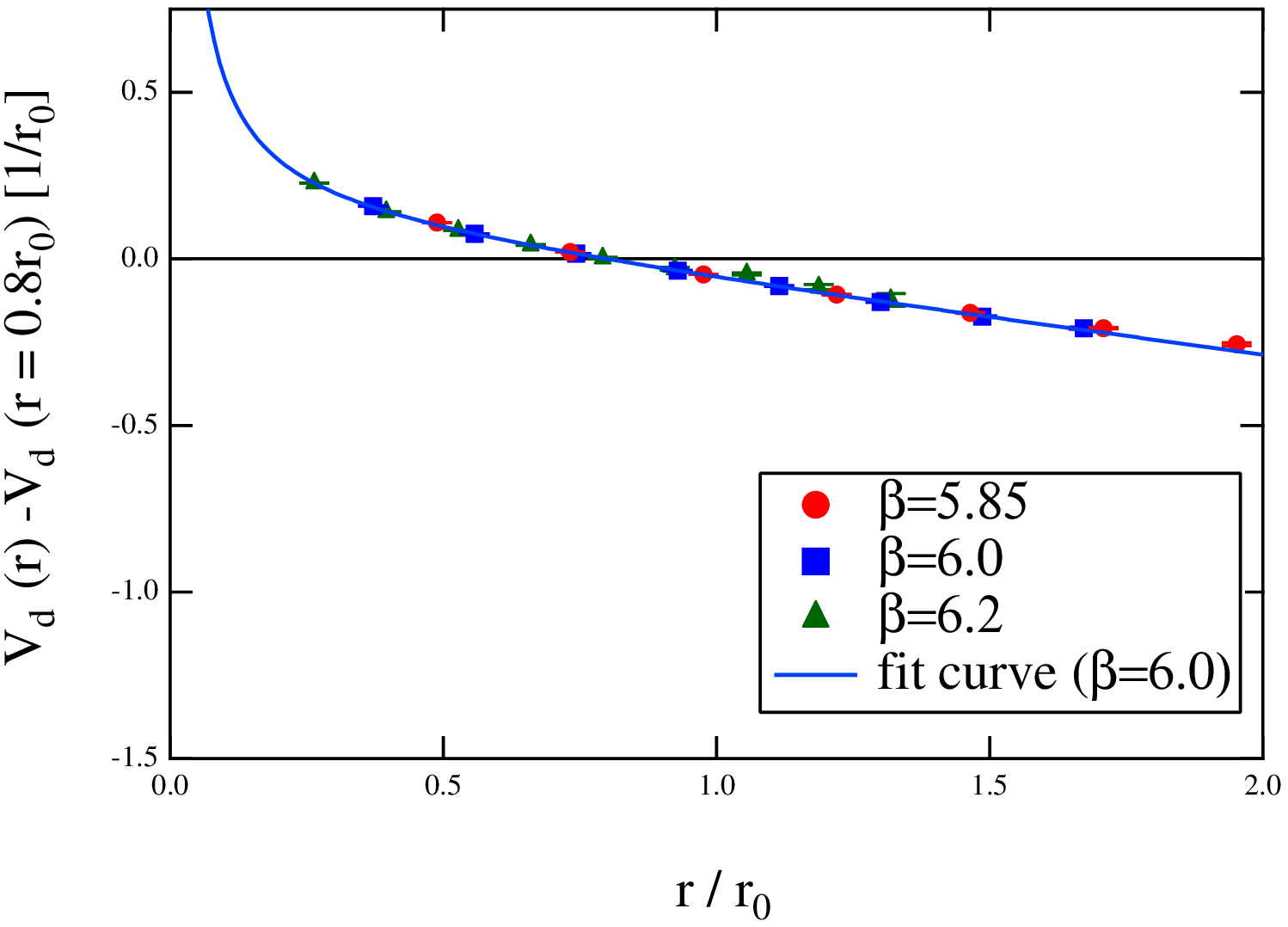}  
\includegraphics[width=\figw]{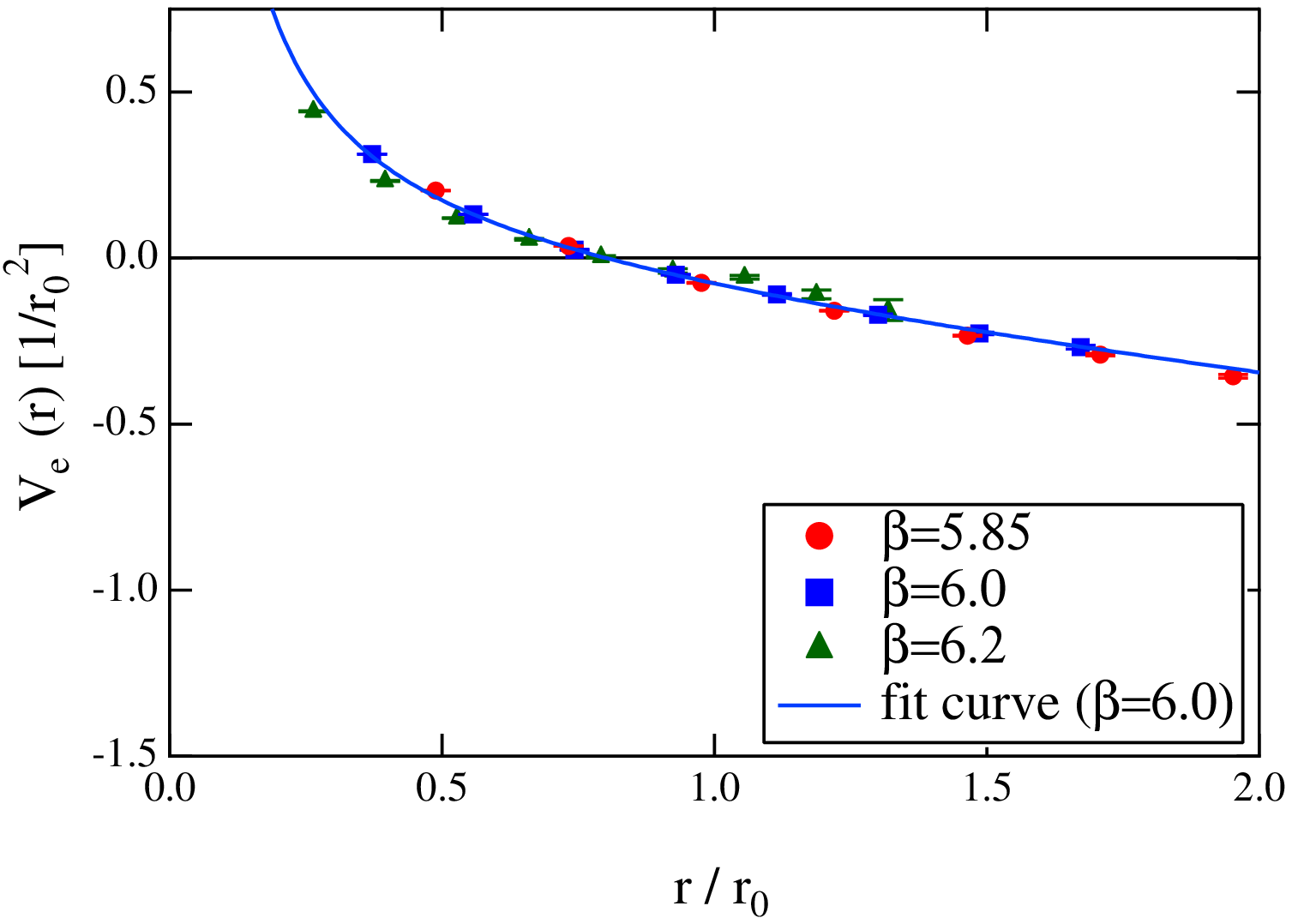}  
\caption{The momentum-dependent corrections $V_{b}(r)$, $V_{c}(r)$, $V_{d}(r)$
and $V_{e}(r)$ in units of $r_{0}$, normalized at $r=0.8\, r_{0}$.}
\label{fig:pot-velocity}
\end{figure}

\subsection{The $O(1/m^2)$ spin-dependent corrections}

In Fig.~\ref{fig:pot-spin}, we present the spin-dependent corrections,
$V'_{1}(r)$, $V'_{2}(r)$, $V_{3}(r)$ and $V_{4}(r)$, where the data 
at $\beta=6.0$ and $6.3$ are already published in Ref.~\cite{Koma:2006fw}.
The $O(1/m^{2})$ spin-orbit corrections, $V_{1}'(r)$ and $V_{2}'(r)$, 
are non-vanishing up to $r=2.23\, r_{0}= 1.12$ fm.
The finite tail of $V_{2}'(r)$ is observed in Ref.~\cite{Koma:2006fw},
which are again observed at further longer distances.

The Gromes relation~\cite{Gromes:1983pm}, 
an important analytic relation derived 
from the Lorentz invariance,
\be
V^{(0)\prime}(r)=V_{2}'(r)-V_{1}'(r) \;,
\label{eq:GromesRelation}
\ee
is approximately satisfied 
as shown in Fig.~\ref{fig:Gromes}, where the comparison of the l.h.s 
and the r.h.s of Eq.~\eqref{eq:GromesRelation} as well as the
relative deviation from the relation, 
$1-(V_{2}'-V_{1}')/V^{(0)\prime}$, are  plotted.
If the Gromes relation is satisfied, this quantity 
should be zero at all $r$.
We find that the deviation is 10 to 12~\% at $a=0.123$~fm, 
while 4 to 10 \% at $a=0.093$~fm,
and seems to be smaller at $a=0.068$~fm.
Namely, we find a tendency such that the deviation decreases as~$a\to 0$.

\par
For the spin-spin corrections, 
$V_{3}(r)$ and $V_{4}(r)$, we see that they have no long range contribution.
However a detailed analysis shows that the functional form slightly deviates
from the leading-order perturbative expressions at intermediate distances.

\subsection{The $O(1/m^2)$ momentum-dependent corrections}

In Fig.~\ref{fig:pot-velocity}, we show the momentum-dependent corrections,
$V_{b}(r)$, $V_{c}(r)$, $V_{d}(r)$ and $V_{e}(r)$,
which are normalized at $r=0.8\, r_{0}$.
Motivated by the minimal area law 
model~\cite{Barchielli:1986zs,Barchielli:1988zp}, we fit the data to the 
functional form as the static potential $V(r) = -A/r +B r +C$, where
the data at $r/a=2$ is not taken into account.
We find that the global structure of the data are well described with this function
as seen in Fig.~\ref{fig:pot-velocity}.

There are nonperturbative relations like the Gromes relation for the 
momentum-dependent corrections,  called the BBMP 
relation~\cite{Barchielli:1986zs,Barchielli:1988zp},
\bea
V_{b}(r)+2V_{d}(r) = -\frac{1}{2}V^{(0)}(r)
+\frac{r}{6}\frac{dV^{(0)}(r)}{dr}
\;,\quad 
V_{c}(r)+2V_{e}(r) = -\frac{r}{2}\frac{dV^{(0)}(r)}{dr} \;.
\label{eqn:bbmp}
\eea
We examined these relations in Ref.~\cite{Koma:2007jq} and found
that they seem to be satisfied.
Using the new data set, we are now investigating the BBMP relation carefully, which
will be reported in a future publication.
In any case, it is clear that the momentum-dependent corrections
contain nonperturbative contribution as they are related to the static potential.
The effect of these corrections to the spectroscopy is not yet well examined, so that
it is quite interesting to solve the Schr\"odinger equation with these corrections.

\section{Summary}

We have investigated the relativistic corrections to the
static potential, the $O(1/m)$ correction and the  $O(1/m^{2})$
spin-dependent and momentum-dependent corrections
in SU(3) lattice gauge theory.
These corrections are important  ingredients of pNRQCD
for heavy quarkonium spectroscopy.
By evaluating the color-electric and color-magnetic FSCs
on the PLCF with the multilevel algorithm, 
and exploiting the spectral representation of FSCs,
we have obtained a very clean signal 
for these corrections in the region from 0.25~fm to 1.2~fm.
We have observed long-range nonperturbative contribution to these corrections, which
show a reasonable scaling behavior with respect to the change
of lattice spacing.

\clearpage


\begin{thebibliography}{10}

\bibitem{Brambilla:2000gk}
N.~Brambilla, A.~Pineda, J.~Soto, and A.~Vairo, {\it The QCD potential at
  {$O(1/m)$}},  {\em Phys. Rev.} {\bf D63} (2001) 014023,
  [\href{http://xxx.lanl.gov/abs/hep-ph/0002250}{{\tt hep-ph/0002250}}].

\bibitem{Pineda:2000sz}
A.~Pineda and A.~Vairo, {\it The {QCD} potential at {$O(1/m^2)$}: {Complete}
  spin-dependent and spin-independent result},  {\em Phys. Rev.} {\bf D63}
  (2001) 054007, [\href{http://xxx.lanl.gov/abs/hep-ph/0009145, {Erratum-{\em
  ibid}} {\bf D64}, 039902 (2001)}{{\tt hep-ph/0009145, {Erratum-{\em ibid}}
  {\bf D64}, 039902 (2001)}}].

\bibitem{Barchielli:1986zs}
A.~Barchielli, E.~Montaldi, and G.~M. Prosperi, {\it On a systematic derivation
  of the quark-antiquark potential},  {\em Nucl. Phys.} {\bf B296} (1988) 625,
  [\href{http://xxx.lanl.gov/abs/{Erratum-{\em ibid}} {\bf B303}, 752
  (1988)}{{\tt {Erratum-{\em ibid}} {\bf B303}, 752 (1988)}}].

\bibitem{Barchielli:1988zp}
A.~Barchielli, N.~Brambilla, and G.~M. Prosperi, {\it Relativistic corrections
  to the quark-antiquark potential and the quarkonium spectrum},  {\em Nuovo
  Cim.} {\bf A103} (1990) 59.

\bibitem{Eichten:1979pu}
E.~Eichten and F.~Feinberg, {\it Spin dependent forces in heavy quark systems},
   {\em Phys. Rev. Lett.} {\bf 43} (1979) 1205.

\bibitem{Gromes:1983pm}
D.~Gromes, {\it Relativistic corrections to the long range quark anti-quark
  potential, electric flux tubes, and area law},  {\em Z. Phys.} {\bf C22}
  (1984) 265.

\bibitem{Koma:2006si}
Y.~Koma, M.~Koma, and H.~Wittig, {\it Nonperturbative determination of the
  {QCD} potential at {$O(1/m)$}},  {\em Phys. Rev. Lett.} {\bf 97} (2006)
  122003, [\href{http://xxx.lanl.gov/abs/hep-lat/0607009}{{\tt
  hep-lat/0607009}}].

\bibitem{Koma:2007jq}
Y.~Koma, M.~Koma, and H.~Wittig, {\it {Relativistic corrections to the static
  potential at {$O(1/m)$} and {$O(1/m^2)$}}},  {\em PoS} {\bf LAT2007} (2007)
  111, [\href{http://xxx.lanl.gov/abs/0711.2322}{{\tt 0711.2322}}].


\bibitem{Koma:2005nq}
M.~Koma, Y.~Koma, and H.~Wittig, {\it Determination of the spin-dependent
  potentials with the multi-level algorithm},  {\em PoS} {\bf LAT2005} (2005)
  216, [\href{http://xxx.lanl.gov/abs/hep-lat/0510059}{{\tt hep-lat/0510059}}].

\bibitem{Koma:2006fw}
Y.~Koma and M.~Koma, {\it Spin-dependent potentials from lattice {QCD}},  {\em
  Nucl. Phys.} {\bf B769} (2007) 79,
  [\href{http://xxx.lanl.gov/abs/hep-lat/0609078}{{\tt hep-lat/0609078}}].

\bibitem{Koma:2008zza}
M.~Koma, Y.~Koma, and H.~Wittig, {\it {Determination of the relativistic
  corrections to the static inter-quark potential from lattice {QCD}}},  {\em
  PoS} {\bf CONFINEMENT8} (2008) 105.

\bibitem{Huntley:1986de}
A.~Huntley and C.~Michael, {\it Spin-spin and spin-orbit potentials from
  lattice gauge theory},  {\em Nucl. Phys.} {\bf B286} (1987) 211.

\bibitem{PerezNadal:2008vm}
G.~Perez-Nadal and J.~Soto, {\it {Effective string theory constraints on the
  long distance behavior of the subleading potentials}},  {\em Phys. Rev.} {\bf
  D79} (2009) 114002, [\href{http://xxx.lanl.gov/abs/0811.2762}{{\tt
  0811.2762}}].

\end{thebibliography}

\providecommand{\href}[2]{#2}\begingroup\raggedright

\end{document}